# Improving Performance of Commercially Available AI Products in a Multi-Agent Configuration

Cory Hymel
Research
Crowdbotics
Berkeley, CA

Sida Peng
Research
Microsoft
Redmond, CA

Kevin Xu
Engineering
GitHub
San Francisco, CA

Charath Ranganathan
Engineering
Crowdbotics
Berkeley, CA

*Abstract*—**In recent years, with the rapid advancement of large language models (LLMs), multi-agent systems have become increasingly more capable of practical application. At the same time, the software development industry has had a number of new AI-powered tools developed that improve the software development lifecycle (SDLC). Academically, much attention has been paid to the role of multi-agent systems to the SDLC. And, while single-agent systems have frequently been examined in real-world applications, we have seen comparatively few real-world examples of publicly available commercial tools working together in a multi-agent system with measurable improvements. In this experiment we test context sharing between Crowdbotics PRD AI, a tool for generating software requirements using AI, and GitHub Copilot, an AI pair-programming tool. By sharing business requirements from PRD AI, we improve the code suggestion capabilities of GitHub Copilot by 13.8% and developer task success rate by 24.5% — demonstrating a real-world example of commercially-available AI systems working together with improved outcomes.**

## I. Introduction

The growing field of AI — and more specifically, large language models (LLMs) — has seen impacts across numerous domains [1, 2, 3]. Single agent systems are powered by one language model and will perform all the reasoning, planning, and tool execution on their own [4]. Multi-agent systems, which are systems consisting of multiple autonomous entities having different information and/or diverging interests [5] have shown promise in improving the capabilities of AI models to perform complex tasks [6, 7, 8, 9, 10]. While single agent architectures excel when problems are well-defined and feedback from other agent-personas or the user is not needed, multi-agent architectures tend to thrive more when collaboration and multiple distinct execution paths are required [11], such as software development.

The software development industry has recently seen a significant influx of AI-powered tools designed to enhance the software development lifecycle (SDLC). These tools, ranging from code completion assistants to requirements engineering platforms, promise to boost productivity and streamline workflows. The performance of single-agent, standalone LLM-based code generation tools has been extensively studied using benchmarks [12, 13]. Multi-agent systems have likewise been studied in academia, with a large body of literature on their application in software development [14, 15, 16]. However, thus far, there have been no controlled studies of commercially-available AI systems working together in a multi-agent model.

In this experiment, we tested two commercially available AI tools: Crowdbotics PRD AI and GitHub Copilot, in an experimental, multi-agent setup where software project requirements generated from PRD AI are shared with GitHub Copilots' neighboring tab context model. By having this additional business context, we expected GitHub Copilot's "code suggestion" feature to improve and developers using PRD AI + GitHub Copilot to succeed more frequently.

### A. AI in Requirements Engineering

Requirements engineering (RE) is fundamental to successful software development. RE comprises the systematic approach to defining, documenting, and maintaining requirements throughout a project's lifecycle. As Nuseibeh and Easterbrook [17] note, RE ensures that the final product aligns with user needs and business objectives. Over the years, many AI techniques have been employed to represent and analyze requirements, ranging from knowledge representation and reasoning in the 1980s to the use of natural language (NL) processing, machine learning, and deep learning since the 2000s [18, 19]. The majority of these methods utilizing ML/DL are based on supervised learning, requiring large amounts of labeled training data not readily available in the RE space [20, 21].

LLMs powered by deep learning algorithms and large training corpus offer significant benefits during the RE process [22]. LLMs' ability to access and generate natural language responses has applications across the RE process, from requirements elicitation, specification extraction, and refinement, and generating solution concepts and system architectures [23]. Recent studies [24] have shown that using just a few prompts, LLMs were able to generate better extraction results than existing techniques such as Jdoctor [25] and DocTer [26].

### B. Background: Crowdbotics PRD AI

Crowdbotics PRD AI tool that specializes in generating robust product requirements documents ("PRD") for software projects leveraging LLMs. The platform uses Azure OpenAI's GPT 4.0 model with a progressive composition model with a retrieval augmented generation (RAG) framework on top of it. The tool is capable of producing extensive requirements that include artifacts such as: epics, user stories, user personas, acceptance criteria, technical recommendations, and others. We used PRD AI to generate a PRD and retrieved a small task list that was used by participants as task requirements and by the GitHub Copilot model for context seeding.

### C. AI in Code Generation

Design and writing software code is something traditionally reserved for humans given the complexity and size of context needed to perform adequately. Recently LLMs have shown amazing abilities in code generation when given a distinctive task with adequate requirements defined [12, 27, 28, 29]. We've seen a number of ways in which code generation powered by LLMs can be used from end-to-end code generation, test generation, snippet generation, code suggestion generation, code completion generation, and others. Research in both the academia and industry settings foreshadows a significant impact on software engineering by boosting developer productivity with LLMs acting as code generators [30, 31, 32, 33].

### D. Background: GitHub Copilot

GitHub Copilot (GHC) is a pair programming tool that has multiple features such as inline suggestions, chat capabilities, and others. GHC is powered by a distinct production version of OpenAI's generative AI model, Codex [12, 34]. GHC itself has been shown to give a large productivity boost to developers, in some cases completing tasks up to 55.8% faster than those not using it [34]. During this study, we focused on measuring only the inline code suggestion functionality of GHC, which at the time of writing has, on average, a 27% acceptance rate [35] (*Fig 1*).

### E. Background: Multi-Agent Systems

In the field of AI, an agent often refers to an artificial entity that is able to perceive its "surroundings", make autonomous decisions, and take actions based on those decisions [36]. The concept and study of "Agent", "Autonomous Agent", "AI Agent", and "Multi-Agent" has been ongoing for decades [37, 38, 39, 40, 41, 42]. Agents can be represented and used in many different ways, from chatbots and copilots to complex autonomous systems [43]. LLMs, with their remarkable ability to reason and non-deterministic design, have made them ideal candidates for creating multi-agent-based systems. Compared to systems using a single LLM-powered agent, multi-agent systems offer advanced capabilities by 1) specializing LLMs into various distinct agents, each with different capabilities, and 2) enabling interactions among these diverse agents to simulate complex real-world environments effectively [44]. This ability to perform complex problem-solving across multiple disciplines has made LLM multi-agent systems attractive adaptations to the software development lifecycle [45, 46].

### F. Background: Multi-Agent Systems in Software Development

Recent research has explored the application of multi-agent systems in software development, leveraging multiple, specialized large language models (LLMs) to improve coherence and correctness in software engineering tasks compared to single-agent systems. MetaGPT introduces a framework that incorporates human workflows and standardized operating procedures into LLM-based multi-agent collaborations, addressing the challenge of cascading hallucinations in complex tasks [45]. Similarly, ChatDev presents a virtual software development company that utilizes LLMs throughout the entire development process, dividing it into distinct stages with specialized software agents [46]. More recently, in the commercial space, Devin was released and claimed as the world's first "AI programmer" [47] however, its ability to handle large projects has yet to be empirically supported.

## II. RESEARCH CONTRIBUTION

While there has been extensive research in academia on the application of multi-agent systems and commercially available "black box" multi-agent systems, there is little to no data on real-world, commercially available, independent models working together to increase their capabilities in a multi-agent framework. This research study was contextualized from Peng et. al's [34] research that demonstrated GitHub Copilot can increase developer productivity by 55.8% on average, in conjunction with Vaithilingam et al.'s [13] findings that Copilot's suggestions sometimes lack alignment with the specific project requirements. By providing GitHub Copilot, an AI code generation tool, with additional requirements context through Crowdbotics PRD AI, an AI requirements generation tool, we demonstrate commercially available products working as a multi-agent system with improved overall performance together.

## III. EXPERIMENT DESIGN

We conducted a controlled experiment to measure the change code suggestion acceptance rate of GHC by developers when performing a coding task by sharing business requirements from PRD AI. The experiment began on August 6, 2024, and concluded on October 11, 2024 — with 101 developers participating.

### A. Participant Selection and Grouping

Participants were sourced from three software development providers: Nagarro, Tkxel, and Upwork. Upon agreement, contracts and data-sharing agreements were sent to participants following vendor policies. In general, the following baseline requirements were provided when searching for possible participant: (*Location*) Global, (*Programming Language*) Python, (*Skill Level*) Intermediate to Advanced, (*Additional Constraint*) Knowledge of Fast API.

We chose to include a global pool of talent to ensure diverse backgrounds and expand the possible participant pool as much as possible. Python was chosen for two reasons. First, Python is a widely known language — making the participant pool larger. Second, Codex, which powers GitHub Copilot, has shown excellent capabilities in Python programming tasks [12]. Participant skill level was self-selected to be between intermediate and advanced with participants all currently working in the software development space.

Finally, we leveraged Fast API [48] as a pre-installed tool to ensure candidates were able to complete the programming task within the 4-hour timeframe where existing knowledge of the API was considered. Upon contract completion, participants were randomly split into three groups with each group being provided unique instructions. The three groups were defined as follows:

**Group 1 (Control Group):** Developers will use VS Code without GitHub Copilot and with starter requirements located in a separate document.

**Group 2 (Copilot Group):** Developers will use VS Code with GitHub Copilot enabled, and with starter requirements located in a separate document.

**Group 3 (Copilot Enhanced Group):** Developers will use VS Code with GitHub Copilot enabled, and with starter requirements pre-seeded in a neighboring tab within VS Code.

### B. Task Definition

The project task was designed to be of moderate difficulty to ensure it was challenging, yet still achievable in the maximum time frame of 4 hours. Participants were compensated based on completion and not by total hours therefore incentivizing them to complete the assignment as quickly as possible [49]. A brief description of the task is listed below. The full description given to participants can be found in the Appendix.

---

**Task Description**:

You are being asked to develop a backend using FastAPI for a (simplified) magazine subscription service. This backend service would expose a REST API that enables users to:

1. Register, login, and reset their passwords.
2. Retrieve a list of magazines available for subscription. This list should include the plans available for each magazine and the discount offered for each plan.
3. Create a subscription for a magazine.
4. Retrieve, modify, and delete their subscriptions.

---

### C. Experiment Procedure and Data Collection

Participants were given a set amount of time (4 hours max) to complete the task to ensure consistency across all three groups. Experiment user flow:

1. **Introduction and Onboarding**: Participants are given access to a different Notion page according to which group they were assigned, which gave them directions to set up their project — along with FAQs, rules, and methods to submit their code upon completion.

2. **Environment Setup**: We used a combination of GitHub Classroom and GitHub Codespaces to preconfigure a web-based IDE environment per group. Participants were given a link depending on which group they were assigned which took them to GitHub Classroom (*Fig 2*). From there, a new repository was automatically generated with pre-seeded "boilerplate" code. This boilerplate code was identical across all the three groups. Next, they were instructed to create a new Codespace, which generates a virtual IDE in their browser (*Fig 3*).

3. **Task Execution**: Participants performed the assigned tasks within the 4-hour timeframe. Participants were asked to ensure they had a complete 4-hour working block available to complete the task.

4. **Task Completion**: When done, participants were instructed to run pre-provide local unit tests (*Fig 4*) to ensure the completeness of the task. After this was completed, they were instructed to submit and push their code to the repository.

5. **Task Validation**: Upon submission to the repository, an automated compiler with test cases checked the validity of the work and ensured completion (Fig 5).

Telemetry data was tracked by GitHub Copilot and reported for study data analysis. The following quantitative data metrics were reported:

| Data Point | Description |
| --- | --- |
| task_acceptance | The number of participants that successfully completed the task in the 4-hour timeframe and were able to pass unit tests. |
| suggestion_acceptance | The percentage of code suggestions from GHC that were accepted by participants |

Fig. 5. Telemetry data points collected for study.

### IV. RESULTS

A total of 101 participants took part in the study, of which 99 participated; 2 became unresponsive shortly after signing the contract and were dismissed as per provider policies. The remaining participants were split into three groups of the following composition: Control Group 32, Copilot Group 35, and Copilot Enhanced Group 32.

### A. Task Acceptance

Of the participants in the Control Group, 11 failed to complete the task. In the Copilot Group, 10 failed to complete the task. In the Copilot Enhanced Group, 3 failed to complete the task. Using linear regression modeling, we found that the Copilot Enhanced group was 24.5% more likely to pass the test than the Control group. The Copilot Enhanced group was also 14.8% more likely to pass the test than participants in the Copilot Group.

|  | **Estimate** | **SE** | **tStat** | **pValue** |
| --- | --- | --- | --- | --- |
| **(Intercept)** | 0.62963 | 0.090556 | 6.9529 | 3.5159e-09 |
| **Treatment_1** | 0.097643 | 0.12211 | 0.79966 | 0.42717 |

Fig. 6. Copilot vs Control, Test Passed

|  | **Estimate** | **SE** | **tStat** | **pValue** |
| --- | --- | --- | --- | --- |
| **(Intercept)** | 0.62963 | 0.079783 | 7.8918 | 1.0399e-10 |
| **Treatment_1** | 0.24537 | 0.10833 | 2.265 | 0.027335 |

Fig. 7. Copilot Enhanced vs Control, Test passed

### B. Suggestion Acceptance

In measuring code suggestions rates, we measured the amount of suggested code provided inline by GitHub Copilot as a ratio of suggestion shown to those accepted. Our hypothesis was that sharing business model context from PRD AI in a neighboring tab with GitHub Copilot would improve the code suggestion rate. This hypothesis was found to be true — using PRD AI with Github CoPilot improved the code suggestion acceptance rate by 13.8 percentage points—from 27% to 40.8%.

|  | Estimate | SE | tStat | pValue |
|---|---|---|---|---|
| (Intercept) | 0.23657 | 0.051156 | 4.6246 | 1.9157e-05 |
| Treatment_1 | 0.13882 | 0.072908 | 1.904 | 0.061475 |

Fig. 8. Copilot fraction acceptance

This offers a substantial improvement (+ 51.1%) in the base code suggestion acceptance rate (27%) of GitHub Copilot users [35]. GHC's code suggestion model uses data from open tabs in an IDE, while these tabs usually contain code, in this experiment we included business requirements that the neighboring tab model could index against resulting in the suggestion improvements. This simple study shows that model context sharing can have a significant impact on commercially available platforms.

There was also a difference in the number of code suggestions shown between the GitHub Group and GitHub Enhanced group. With the GitHub Group being shown 257.06 on average and the GitHub Enhanced Group being shown 127.09 on average. While there were fewer lines shown the GitHub Enhanced Group, they were of higher quality given the increased acceptance rate. It's also reasonable to suggest that there were fewer suggestions shown because participants in the GitHub Enhanced Group arrived at the natural stopping point of "this code works" in fewer prompts.

### C. Study Limitations

This study's focus on a single, specific coding task within a 4-hour time limit, while necessary for experimental control, may not fully reflect the diversity and extended duration of real-world development scenarios. Additionally, individual variations in programming expertise and prior experience with AI coding assistants could have influenced the results, despite our efforts to control for skill level. The study also assumes that the context provided by PRD AI was uniformly beneficial, which may not always be the case in real-world scenarios where context quality and relevance can vary. Lastly, our research does not address the long-term effects of using such integrated AI tools on developer skills and practices.

## V. Discussion

Our findings in this paper show the potential opportunity for commercial model enhancement in a multi-agent setup. The combination of PRD AI's business context with GitHub Copilot led to a substantial 13.8% increase in code suggestion acceptance rates. This improvement is particularly significant, as it represents a 51.1% increase over the baseline acceptance rate of 27% for typical GitHub Copilot users. This significant improvement in code suggestion acceptance rates underscores the importance of context in AI-assisted coding, suggesting that future AI software development tools could benefit greatly from incorporating broader project context through other tooling providers.

Participants using the Copilot Enhanced setup were 24.5% more likely to complete the assigned task successfully compared to the Control Group, and 14.8% more likely to complete the task compared to the GitHub Copilot. These results indicate that context-enhanced AI tools could lead to substantial productivity gains in real-world software development scenarios. In fact, the overall candidate success rate in this experiment shows the significant benefit of creating commercially-available AI tooling that is capable of integrating and sharing with other providers.

This study demonstrates the significant potential of context-sharing between commercially available AI tools to enhance their overall performance. As outlined in the Introduction, a significant amount of research has focused on single-agent architecture, with new work in multi-agent configurations constrained to in either academia or closed systems. We have yet to see commercially-available platforms openly work to create accessible, model-to-model collaborative systems. Based on our findings in this study, we hope to promote a more open, collaborative AI future.

### Acknowledgment

We would like to thank Nagarro on their close partnership in helping identify, and manage study participant's quickly. Darcy Jacobsen from The Wednesday Group for continuous proofing and editing before publication. All the amazing developers that participated in the study, thank you.

APPENDIX

```
def calculateDaysBetweenDates(begin, end) {
    var date1 = new Date(begin);
    var date2 = new Date(end);
    var diff = Math.abs(date2 - date1);
    var days = Math.ceil(diff / (1000 * 60 * 60 * 24));
    return days;
}
```

*Fig. 1. Example of code suggestions from GitHub Copilot*

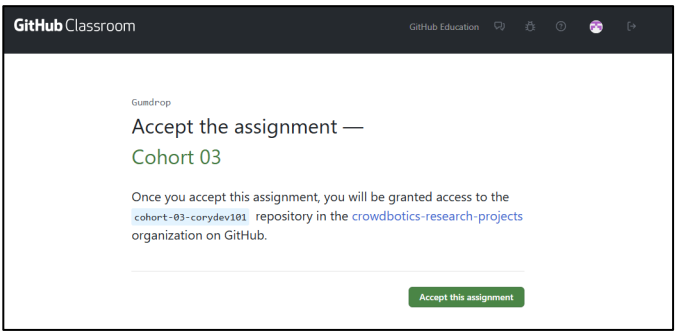

*Fig. 2. GitHub Classroom particpants accessed*

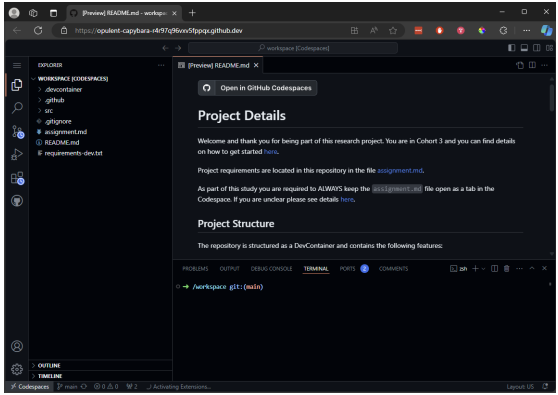

*Fig. 3. Browser based IDE, Codespace, from GitHub.*

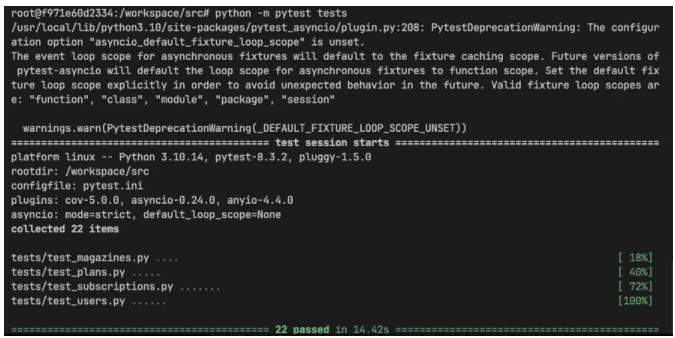

*Fig. 4. Example of local tests passing*

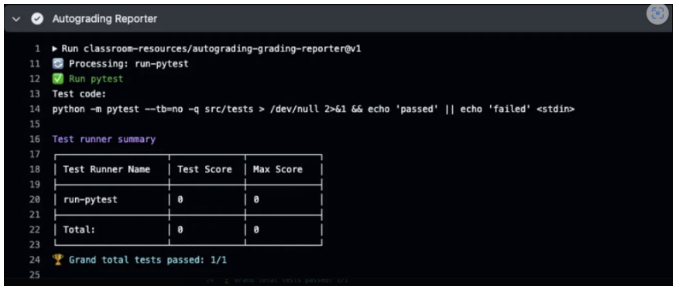

*Fig. 5. Example Autograder test passing*

## I. Appendix: Project Requirements

# Requirements for Magazine Subscription Service

You are being asked to develop a backend using FastAPI for a (simplified) magazine subscription service. This backend service would expose a REST API that enables users to:

1. Register, login, and reset their passwords.
2. Retrieve a list of magazines available for subscription. This list should include the plans available for each magazine and the discount offered for each plan.
3. Create a subscription for a magazine.
4. Retrieve, modify, and delete their subscriptions.

## Data Models Overview

### Magazine

A magazine that is available for subscription. Includes metadata about the magazine such as the name, description, and a base_price (which is the price charged for a monthly subscription). The base_price is a numerical value and must be greater than zero.

### Plan

Plans to which users can subscribe their magazines. There are 4 plans available in the system as described below.

A Plan object has the following properties: title, a description, a renewalPeriod, discount - a percentage, expressed as a decimal - for this plan (e.g. a discount of 0.1 means a 10% discount), and a tier. The tier is a numerical value that represents the level of the plan. The higher the tier, the more expensive the plan.

The renewalPeriod is a numerical value that represents the number of months in which the subscription would renew. Renewal periods CANNOT be zero. For example, a renewalPeriod of 3 means that the subscription renews every 3 months.

The 4 plans that you must support are given below.

#### Silver Plan

- title: "Silver Plan"
- description: "Basic plan which renews monthly"
- renewalPeriod: 1
- tier: 1
- discount: 0.0

#### Gold Plan

- title: "Gold Plan"
- description: "Standard plan which renews every 3 months"
- renewalPeriod: 3
- tier: 2
- discount: 0.05

#### Platinum Plan

- title: "Platinum Plan"
- description: "Premium plan which renews every 6 months"
- renewalPeriod: 6
- tier: 3
- discount: 0.10

#### Diamond Plan

- title: "Diamond Plan"
- description: "Exclusive plan which renews annually"
- renewalPeriod: 12
- tier: 4
- discount: 0.25

### Subscription

A Subscription tracks which Plan is associated with which Magazine for a specific User. The subscription also tracks the price at renewal for that magazine and the next renewal date.

A User can have only one Subscription for a specific Magazine and Plan at a time. The Subscription object has the following properties: user_id, magazine_id, plan_id, price, renewal_date, and is_active.

The price at renewal is calculated as the base_price of the magazine discounted by the discount of the plan. For example, if the base price of the magazine is 100 and the plan discount is 0.10, the price will be 90. The price is a numerical value and must be greater than zero.

For record keeping purposes, subscriptions are never deleted. If a user cancels a subscription to a magazine, the corresponding is_active attribute of that Subscription is set to False. Inactive subscriptions are never returned in the response when the user queries their subscriptions.

## Business Rules

1. Subscriptions can be modified before the expiry of the subscription period. For example, if a user has subscribed to a magazine with a Silver Plan and decides to upgrade to a Gold Plan, the Silver Plan subscription is deactivated and a new subscription is created with a new renewal date for the Gold Plan that the user has chosen.
2. If a user modifies their subscription for a magazine, the corresponding subsciption is deactivated and a new subscription is created with a new renewal date depending on the plan that is chosen by the user.
1. For this purpose assume that there is no proration of the funds and no refunds are issued.